\documentclass[twocolumn,prl,showpacs,amsmath]{revtex4}
\usepackage{epsfig,amsfonts,amsbsy}
\newcommand{\bra}{\left\langle}
\newcommand{\ket}{\right\rangle}
\newcommand{\pder}[2]{\frac{\partial #1}{\partial  #2}}
\newcommand{\pderf}[3]{\left(\frac{\partial #1}{\partial  #2}\right)_{#3}}

\newcommand{\der}[2]{\frac{d #1}{d  #2}}

\newcommand{\bv}[1]{{\boldsymbol #1}}

\newcommand{\ep}{\epsilon}
\newcommand{\cI}{{\cal I}}
\begin{document}
\title{Thermodynamic entropy as a Noether invariant}
\author{Shin-ichi Sasa}
\affiliation {
Department of Physics, Kyoto University, Kyoto 606-8502, Japan}
\author{Yuki Yokokura}
\affiliation {International Centre for Theoretical Sciences, 
Survey No.151, Shivakote, Hesaraghatta Hobli, 
Bengaluru North - 560 089, India.}


\begin{abstract}
We study a classical many-particle system with an external control 
represented by a time-dependent extensive parameter in a Lagrangian. 
We show that thermodynamic entropy of the system is uniquely 
characterized as the  
Noether invariant associated with a symmetry 
for an infinitesimal non-uniform time translation $t\to t+\eta\hbar \beta$, 
where $\eta$ is a small parameter, $\hbar$ is the Planck constant, 
$\beta$ is the inverse temperature 
that depends on the energy and control parameter, and 
trajectories in the phase space are restricted to 
those consistent with quasi-static processes in thermodynamics.
\end{abstract}

\maketitle

{\em Introduction.---}
Entropy is a fundamental concept in physics. It appears in 
thermodynamics \cite{Thermodynamics,Lieb-Yngvason},
statistical mechanics \cite{Stat-Mech},
information theory \cite{information},
computation theory  \cite{complexity},
quantum information theory \cite{entanglement},
and thermodynamics of black holes \cite{black-hole}. 
Recently, the inter-relation between different types of entropy
has been discovered. The second law of thermodynamics has 
been extended so as to apply systems with a feedback control 
through exchange of information, not of energy, 
between the system and the controller \cite{Sagawa}. This opens 
up studies in the intersection of thermodynamics and 
information theory \cite{information-thermodynamics}. 
As another development, there have been attempts to 
connect black hole entropy to entanglement entropy 
\cite{BH-entangle-1,BH-entangle-2}, 
and in the AdS/CFT context a novel notion of holographic entanglement entropy 
has appeared, which provides a dual description between boundary entanglement 
entropy and dynamics of bulk spacetime \cite{RT}. 
By synthesizing various aspects of entropy, 
we thus obtain a deeper understanding of fundamental laws in physics. 
Now, 
there is a paper {\it``Black hole entropy is the Noether charge''} \cite{Wald},
which claims that black hole entropy is obtained as the 
Noether charge associated with the horizon Killing field. 
We are then naturally led to ask whether thermodynamic entropy of
standard materials is also characterized by a Noether invariant.


Suppose that  we have a many-particle isolated system in a box, 
and that an external controller moves a piston, 
which may be described by a time-dependent single-body potential. 
Then, in response to the fact that thermodynamic 
entropy keeps a constant value 
in quasi-static adiabatic processes \cite{Thermodynamics}, 
it was proved that
along almost all the solution trajectories to the equation of motion 
with quasi-static change in the volume, 
the phase space volume enclosed by the energy surface including 
the phase space point at time $t$ is invariant \cite{Anosov,Kasuga, 
Ott,Lochak-Meunier,Jarzynski93,Berry-Robbins}. 
Thus, the logarithm of the phase space volume provides a definition 
of time-dependent entropy in mechanics. 
The main result of this Letter is that 
there exists a symmetry by which the entropy is uniquely
characterized as a Noether invariant. 


The key step in our theory is to formulate a special class of 
trajectories that are consistent with quasi-static processes in thermodynamics. 
By restricting the domain of the action to this class of trajectories,
we find a symmetry for an infinitesimal non-uniform time 
translation $t\to t+\eta\hbar \beta$, 
where $\eta$ is a small parameter, $\hbar$ is the Planck constant, and  $\beta $ 
is the inverse temperature determined 
by applying the thermodynamic relation to the time-dependent entropy. 
It should be noted that our theory stands on classical mechanics, 
classical statistical mechanics, and thermodynamics;
and thus the Planck constant does not appear. 
Nevertheless, our theory leads to the existence of a universal constant 
with the same dimension as the action. 


Below, we first describe a setting up of classical mechanics
of the particle system, and discuss a generalized Noether 
theorem associated with a symmetry. We then define trajectories
consistent with quasi-static processes based on statistical 
mechanics. By combining these two concepts, we derive our main result. 

{\em Mechanics.---}
Let $q(t) \in {\mathbb{R}}^{3N}$ be a collection of coordinates of 
$N$ particles with short-range interaction in a box of volume $V$.  
We particularly focus on  macroscopic systems where the extensive 
behavior is observed for large $N$. We denote the trajectory 
$(q(t))_{t=t_{\rm i}}^{t_{\rm f}}$ by $\hat q$. 
We also introduce an extensive control parameter $\alpha$, 
whose typical example is the volume $V$. 
(Formally, $\alpha$ is a complete set of extensive work variables.) 
For a fixed protocol of the parameter 
$\hat \alpha =(\alpha(t))_{t=t_{\rm i}}^{t_{\rm f}}$,
the action $\cI(\hat q,\hat \alpha)$ is given by 
\begin{equation}\label{Action}
\cI(\hat q,\hat \alpha) = 
\int_{t_{\rm i}}^{t_{\rm f}} dt L(q(t), \dot q(t), \alpha(t)),
\end{equation}
where the dot denotes the time derivative. 
All the mechanical properties are represented by the Lagrangian \cite{SM-d6}. 
We also assume that there is no conserved quantity other than the total
energy for the system with $\alpha$ fixed, 
$E(q,\dot q,\alpha) = \dot q \partial {L}/\partial {\dot q} -  L(q,\dot q,\alpha)$. 

We consider a non-uniform time translation: 
$t \to  t'=t+\eta \xi(q,\dot q, \alpha)$. 
Here $\eta$ is a small parameter, and the functional form of 
$\xi$ is not specified yet. 
Then, the transformation $\hat q \to \hat q'$ is given by $q'(t')=q(t)$, 
because the position of particles is independent of relabeling time coordinate. 
The transformation $\hat\alpha \to \hat\alpha'$ corresponds to 
$\alpha'(t')=\alpha(t')$, 
because the protocol $\hat\alpha$ is fixed. 
We represent this transformation by index $G$, and neglect the 
contribution of $O(\eta^2)$. 
Then, the change in action 
$
\delta_G \cI  \equiv \cI(\hat q',\hat \alpha')- \cI(\hat q,\hat \alpha)
$
is expressed as 
\begin{equation}
\delta_G \cI  = 
\int_{t_{\rm i}}^{t_{\rm f}} dt  
\left[
\bar \delta_G L + \eta \der{(\xi L)}{t} \right],
\end{equation}
where we have defined 
$\bar \delta_G L \equiv L(q'(t), \dot q'(t), \alpha'(t))-
L(q(t), \dot q(t), \alpha(t))$. 
Noting that $\bar \delta_G q(t)\equiv q'(t)-q(t)=-\eta \xi \dot q$ 
and introducing the Euler-Lagrange derivative
\begin{equation}
{\cal E} \equiv  \pder{L}{q}-\der{}{t} \pder{L}{\dot q},
\end{equation}
we express $\bar \delta_G L$ in terms of $\bar \delta_G q$. 
Thus, we obtain 
\begin{equation}
\delta_G \cI  = 
\eta \int_{t_{\rm i}}^{t_{\rm f}} dt \left\{ - {\cal E} \dot q \xi 
+  \der{}{t} \left[\xi \left(L- \dot q \frac{\partial {L}}{\partial {\dot q}}\right) \right] \right\}.
\label{form-1}
\end{equation}

Now suppose that, for some $\hat \alpha$,  
there exist $\xi(q,\dot q, \alpha)$ and $\psi(q,\dot q,\alpha) $ such that
\cite{Trautmen} \cite{fn:Noether} 
\begin{equation}
\delta_G \cI= \eta \int_{t_{\rm i}}^{t_{\rm f}} dt \der{\psi}{t}
\label{cartan-2}
\end{equation}
for a class of trajectories $\hat q$, which is identified later. 
Then, (\ref{form-1}) is written as 
\begin{equation}
\int_{t_{\rm i}}^{t_{\rm f}} dt {\cal E} \dot q \xi 
 = - (\psi+E \xi) |_{t_{\rm i}}^{t_{\rm f}}.
\label{form-11}
\end{equation}
This leads to two important properties.
First, because ${\cal E}=0$ at any solution 
$\hat q_*$, 
we obtain a conservation law
\begin{equation}
(\psi_*+  E_* \xi_*) |_{t_{\rm i}}^{t_{\rm f}}=0.
\label{conserve-2}
\end{equation}
Here, the subscript of $B_*$ represents the evaluation of
a quantity $B$ at a solution trajectory $q_*(t)$. 
Second, by substituting $q(t)= q_*(t+\eta \xi_*)$ into
(\ref{form-11}), we have 
\begin{equation}
\left. \int_{t_{\rm i}}^{t_{\rm f}} dt {\cal E} \dot q \xi \right\vert_{q=q_*(t+\eta\xi_*)} 
 = - \left. (E \xi+\psi) |_{t_{\rm i}'}^{t_{\rm f}'} \right\vert_*,
\label{form-12}
\end{equation}
where we have used $q_*(t_{\rm i}+\eta \xi_*(t_{\rm i}))=q_*(t_{\rm i}')$.  
Because the conservation law (\ref{conserve-2}) holds for 
any $t_{\rm i}$ and $t_{\rm f}$, 
the right-hand side of (\ref{form-12}) is equal 
to zero. Expanding the left-hand side with respect to $\eta$, we obtain
\begin{equation}
\int_{t_{\rm i}}^{t_{\rm f}} dt \left. \frac{\delta {\cal E}}{\delta q } (\bar \delta_G q)
\dot q \xi  \right|_* = 0,
\label{form-13}
\end{equation}
where we have used the equation of motion ${\cal E}|_*=0$. 
The relation (\ref{form-13}) implies that $q_*+\bar \delta_G q|_*$ 
is  a solution of the same equation of motion \cite{Trautmen}. 
That is, the transformation $G$ 
maps each solution trajectory to another one in 
the system $\cI(\hat q,\hat \alpha)$. 
This property was referred to
as a {\it dynamical symmetry} \cite{Prince,SIAM}. 
If $\psi$ in (\ref{cartan-2}) is independent of $\dot q$, 
which includes the case $\psi=0$, 
$\cI(\hat q',\hat \alpha')$ provides the same equation of motion as that 
for $\cI(\hat q,\hat \alpha)$. 
In a more general case where $\psi$ depends on $\dot q$, 
the action $\cI(\hat q',\hat \alpha')$ defines a different dynamical system. 
Even for this case, however, 
(\ref{cartan-2}) represents a symmetry, 
leading to the dynamical symmetry 
and the conservation law (\ref{conserve-2}), as we have seen above. 
This was called a generalized Noether theorem  \cite{fn:symmetry}. 
In this context, 
$\psi+E \xi$ is the Noether invariant associated with the transformation $G$.

{\em Thermodynamics.---}
Let us briefly review statistical mechanics. 
We introduce a phase space coordinate $\Gamma=(q,p)$ with
the momentum $p \equiv  \partial{L}/\partial {\dot q} \in {\mathbb{R}}^{3N}$, 
and assume that $\dot q$ can be uniquely determined for $(q,p)$. Then, 
$H(\Gamma,\alpha)= E(q,\dot q(q,p),\alpha)$ is  the Hamiltonian.
The expectation of any quantity $A(\Gamma)$ with 
respect to the micro-canonical ensemble of $(E,\alpha)$ is defined as 
\begin{equation}
\bra A \ket^{\rm mc}_{E,\alpha}
\equiv  
\frac{1}{\Sigma(E,\alpha)} 
\int d \Gamma \delta(E-H(\Gamma,\alpha)) A(\Gamma),
\end{equation}
where $\Sigma(E,\alpha)  \equiv  \int d\Gamma \delta(E-H(\Gamma,\alpha))$ 
is the normalization constant. 
Throughout this Letter, the Boltzmann constant is set to unity.
According to the formula in statistical mechanics, the entropy $S$ 
is defined as 
\begin{equation}
S(E,\alpha)\equiv \log \frac{\Omega(E,\alpha)}{N!}
\label{S-formula}
\end{equation}
with 
$
\Omega(E, \alpha) \equiv  \int d\Gamma \theta(E-H(\Gamma,\alpha)),
$
where $\theta(x)=1$ for $x\ge 0$ and $\theta(x)=0$ for $x < 0$ \cite{S_hbar}.
We can then confirm the fundamental relation in thermodynamics \cite{SM-d0}:
\begin{equation}
dS=\beta dE - \beta\bra \pder{H}{\alpha} \ket_{E,\alpha}^{\rm mc} d\alpha 
\label{f-relation}
\end{equation}
with the definition of the inverse temperature 
\begin{equation}
\beta \equiv  \frac{\Sigma(E,\alpha)}{\Omega(E,\alpha)}.
\end{equation}
When $\alpha$ represents the volume $V$, 
the second term of the right-hand side of (\ref{f-relation}) 
becomes $\beta PdV$ with the pressure 
$P=-\bra \partial H/\partial V \ket^{\rm mc}_{E,\alpha}$. 
In general, the relation (\ref{f-relation}) guarantees  
the consistency with thermodynamics. 


In the following argument, we consider the quasi-static change in $\alpha$. 
This is realized by choosing $\alpha(t)= \bar \alpha (\epsilon t)$, 
where the functional form of $\bar \alpha$ is independent of $\epsilon$, 
introducing $\tau = \epsilon t$ and  taking the {\it quasi-static
limit } $\epsilon \to 0$ with $\tau_{\rm i} = \epsilon t_{\rm i}$ 
and $\tau_{\rm f}=\epsilon t_{\rm f}$ fixed. 
Indeed, $d\alpha/dt=\epsilon d\bar\alpha/d\tau=O(\epsilon)$. 
Now, we take a solution trajectory $\Gamma_*(t)$, 
which is realized in the ideally isolated mechanical system. 
Then, it determines the time evolution of the energy as 
$E_*(t)=H(\Gamma_*(t),\alpha(t))$. As the result, 
the time evolution of the entropy and inverse temperature 
is also obtained by $S(E_*(t),\alpha(t))$ and 
$\beta(E_*(t),\alpha(t))$, respectively. 
The adiabatic theorem tells us that 
$S(E_*(t),\alpha(t))$ keeps a constant value  along 
almost all solution trajectories 
in the quasi-static limit \cite{Anosov,Kasuga,Ott,Lochak-Meunier,
  Jarzynski93,Berry-Robbins,SM-d2}. 
This means that in the quasi-static limit, 
almost all solution trajectories with the same initial energy 
give the same {\it adiabatic curve}
in the thermodynamic state space $(E, \alpha)$. 
On the basis of the ideally isolated mechanical system, 
thus we have a mechanical description consistent with thermodynamics.

Let us now consider a more realistic situation in which 
our $N$-particle system enclosed by adiabatic walls is 
not completely isolated. 
Then, trajectories of the particles are not solutions 
to the equation of motion for the Lagrangian (\ref{Action}), 
because the constituents of the walls may influence the motion 
of the particles. 
Even for this case, however, it can be assumed ideally 
that the $N$-particle system is thermally isolated (which means  adiabatic in 
thermodynamics) and that the 
entropy keeps a constant value in quasi-static processes. 
Motivated by this fact, 
we try to characterize such phase-space trajectories. 


We first identify the condition of phase space trajectories 
consistent with quasi-static processes in thermodynamics, 
which are not necessarily  solution trajectories for
our Lagrangian (\ref{Action}). 
We refer to such trajectories as
{\it thermodynamically consistent trajectories}. 
Suppose a curve $(\bar E(\tau), \bar \alpha(\tau))$,
$\tau_{\rm i} \le \tau \le \tau_{\rm f}$, in the thermodynamic 
state space, which corresponds to a quasi-static process in thermodynamics. 
Here $\bar E(\tau)$ is obtained by $E(t)=\bar E(\epsilon t)$, 
which follows the change of $\bar \alpha(\tau)$. 
Then, 
for thermodynamically consistent trajectories, 
the mechanical work $\int dt
\left({d\alpha}/{dt}\right)
\left(\partial {H}/\partial {\alpha}\right)$
is expected to be equal to the thermodynamic work
$\int dt \left( d\alpha/{dt} \right)
\bra \left(\partial {H}/\partial {\alpha} \right)
\ket_{E(t), \alpha(t)}^{\rm mc}$.
We thus \textit{define} thermodynamically consistent trajectories 
as those satisfying
\begin{equation}
\lim_{\epsilon \to 0}
\int^{\tau_{\rm f}'}_{\tau_{\rm i}'} d\tau \frac{d\bar\alpha}{d \tau}
\left[\pder{H}{\alpha}-\bra \pder{H}{\alpha}
  \ket_{\bar E(\tau), \bar \alpha(\tau)}^{\rm mc}\right]=0
\label{quasi-static}
\end{equation}
for any time interval $[\tau_{\rm i}', \tau_{\rm f}']$ such that 
$\tau_{\rm i} \le \tau_{\rm i}' <  \tau_{\rm f}' \le \tau_{\rm f}$. 
Here, it should be noted that $\partial{H}/\partial{\alpha}$
is a rapidly varying function of $\tau$ because it depends
on $\Gamma(\tau/\epsilon)$  \cite{SM-d11}. 

Next, we determine the adiabatic condition. 
Let us fix an adiabatic curve and consider phase space 
trajectories that yield the adiabatic curve. 
From the expression $E(t)=H(\Gamma(t),\alpha(t))$ 
for any $\Gamma(t)$, we have 
\begin{equation}
\der{E}{t}=\pder{H}{\Gamma}\dot \Gamma+\pder{H}{\alpha}\dot \alpha.
\end{equation}
If the trajectory describes the behavior of a thermally isolated system, 
the energy changes only through the external control. 
This property can be represented by 
\begin{equation}
\pder{H}{\Gamma}\dot \Gamma=0. 
\label{e-con}
\end{equation}
This is the condition of the idealized adiabatic wall, 
which solution trajectories satisfy, of course. 

Finally, we check that 
$S(t_{\rm f})=S(t_{\rm i})$ holds 
for thermodynamically consistent trajectories satisfying (\ref{e-con}). 
Here, $S(t)\equiv S(H(\Gamma(t),\alpha(t)),\alpha(t))$ for (\ref{S-formula}). 
By using (\ref{f-relation}) and noting
that ${d\bar E}/{d\tau} =\left( \partial{H}/\partial {\alpha}\right)
\left({d\bar \alpha}/{d\tau}\right)$
under (\ref{e-con}), 
we express $S(t_{\rm f})-S(t_{\rm i})=\int^{\tau_{\rm f}}_{\tau_{\rm i}}
d\tau  \left( dS(\bar E(\tau),\bar \alpha(\tau))/{d\tau} \right)$ 
as
\begin{equation}
\int_{\tau_{\rm i}}^{\tau_{\rm f}} d\tau \beta  \frac{d\bar \alpha}{d\tau}
\left[\pder{H}{\alpha}-\bra \pder{H}{\alpha} \ket_{\bar E(\tau), \bar\alpha(\tau)}^{\rm mc}\right].
\label{S_inv}
\end{equation}
Because $\beta(\tau)=\beta(\bar E(\tau), \bar\alpha(\tau))$
is a slowly varying function of $\tau$, using 
$\tau_k=(\tau_{\rm f}-\tau_{\rm i})k/K +\tau_{\rm i}$ with large $K$,
(\ref{S_inv}) may be estimated as 
\begin{equation}
  \sum_{k=1}^K \beta(\tau_k) \int_{\tau_{k-1}}^{\tau_{k}}
  d\tau   \frac{d\bar \alpha}{d\tau}
  \left[\pder{H}{\alpha}-\bra \pder{H}{\alpha}
    \ket_{\bar E(\tau), \bar\alpha(\tau)}^{\rm mc}\right]
\label{S_inv-2}
\end{equation}
with an accuracy of $O(1/K)$. 
Then, (\ref{S_inv-2}) tends to zero as $\epsilon \to 0$
due to (\ref{quasi-static}), and (\ref{S_inv}) is estimated
as zero for infinitely large $K$. In the following, this invariance is
expressed by the generalized Noether theorem. 

{\em Main result.---}
We now derive the thermodynamic entropy (\ref{S-formula}) 
as the Noether invariant $\psi+E\xi$ associated with a transformation $G$. 
First, we recall that the symmetry exists 
only if there are $\xi$ and $\psi$ satisfying (\ref{form-11}). 
For the general Lagrangian we study, 
there are no such $\xi$ and $\psi$ for arbitrary $\hat q$ 
and $\hat \alpha$, which is consistent with a fact 
that the entropy is invariant only in quasi-static 
adiabatic processes. When we attempt to understand 
thermodynamic properties, 
we have to  study 
thermodynamically consistent trajectories.
Hence, we can expect that for them 
there exist $\xi$  and $\psi$ satisfying (\ref{form-11}). 
We shall show this from now.  
By using the identity
\begin{equation}
\der{E}{t}= -{\cal E} \dot q  +\pder{E}{\alpha} \dot \alpha,
\label{energy-balance}
\end{equation}
we rewrite (\ref{form-11}) as 
\begin{equation}
\int_{t_{\rm i}}^{t_{\rm f}} dt \xi\left[ \frac{dE}{dt}-\pder{E}{\alpha} \dot \alpha \right]
= \int_{t_{\rm i}}^{t_{\rm f}} dt \frac{d(\psi + \xi E)}{dt}. 
\label{identity23}
\end{equation}
Suppose that $\xi = \Xi(E(q,\dot q, \alpha),\alpha)$ and 
$\psi = \Psi(E(q,\dot q, \alpha),\alpha)$ satisfy (\ref{identity23}).
Then, in the quasi-static limit, (\ref{identity23}) becomes 
\begin{equation}
\int_{\tau_{\rm i}}^{\tau_{\rm f}} d\tau \Xi\left[ \frac{d\bar E}{d\tau}-\bra \pder{H}{\alpha} \ket^{\rm mc}_{\bar E(\tau),\bar \alpha(\tau)}\frac{d\bar \alpha}{d\tau} \right]
= \int_{\tau_{\rm i}}^{\tau_{\rm f}} d\tau \frac{d(\Psi + \Xi \bar E)}{d\tau} 
\label{threorem-con-2}
\end{equation}
for thermodynamically consistent trajectories \cite{SM-d3}. 
When there exist $\Xi$ and $\Psi$ satisfying this equation, 
it should hold for any $\tau_{\rm f}$.
This means that the integrand in (\ref{threorem-con-2}) itself 
vanishes  for each $\tau$, and hence we have
\begin{equation}
\Xi \left[ d\bar E -\bra \pder{H}{\alpha} \ket_{\bar E,\bar \alpha}^{\rm mc} d\bar \alpha\right]
=d(\Psi+\Xi\bar E).
\label{sym-6}
\end{equation}


Let us solve (\ref{sym-6}). 
Because the right-hand side is 
a total derivative of a function of $(E,\alpha)$ \cite{fn:integrating},  
the necessary and sufficient
condition for the existence of $\Psi(E,\alpha)$ in (\ref{sym-6}) is given 
by the integrability condition: 
\begin{equation}
\pderf{\Xi}{\alpha}{E}+\pder{}{E}
\left( \Xi \bra \pder{H}{\alpha} \ket_{E,\alpha}^{\rm mc} \right)_{\alpha}=0.
\label{eq-Xi}
\end{equation}
By using  (\ref{f-relation}), we express the left-hand side as
\begin{equation}
\pder{}{\alpha}\left(\Xi \beta^{-1} \pderf{S}{E}{\alpha}\right)_E
- \pder{}{E}
\left( \Xi \beta^{-1} \pderf{S}{\alpha}{E}  \right)_\alpha .
\end{equation}
Then, we find that the functional determinant 
$|\partial(\Xi \beta^{-1},S)/\partial(\alpha,E)|$ vanishes.  
This means  that $\Xi=\beta {\cal F}(S)$, 
where ${\cal F}$ is an arbitrary function of $S$ \cite{SM-d5}. 
By substituting this into (\ref{sym-6}), 
employing (\ref{f-relation}), and integrating it, 
we obtain the Noether invariant: 
\begin{equation}
\Psi+E \Xi =  \int^S dS' {\cal F}(S'). 
\label{integral}
\end{equation}
Note that this is conserved even for thermodynamically consistent adiabatic 
non-solution trajectories 
because the left-hand side of (\ref{form-11}) vanishes 
due to (\ref{e-con}).


In particular, we study  the Noether invariant $\Psi+E\Xi$ described
by an extensive variable for a macroscopic equilibrium
system. In this case, the transformation of $\Psi+E\Xi$ for
size scaling leads to the result that $\Psi$ is extensive and $\Xi$
is intensive. Because
$\beta$ is intensive, $\Xi \beta^{-1}={\cal F}(S)$ becomes
a special intensive variable 
that does not depend explicitly on
the extensive work variable $\alpha$ such as the volume $V$. 


Let us determine the functional form of ${\cal F}(S;{\rm M},N)$,
where we explicitly write the dependence on the type of material ${\rm M}$
and the particle number $N$. The most important property of macroscopic
systems is the additivity. As an example, we consider a composite
system that consists of two macroscopic subsystems A and B in thermal
contact.
In the following, we denote physical quantities $Q$ and the type of
material ${\rm M}$ in the subsystem ${\rm X}$ 
by $Q_{\rm X}$ and ${\rm M}_{\rm X}$, respectively, where X$=$A or B. 


Now, the time translation $t \to t+\eta \Xi$ is applied to
the composite system. Because the time coordinate is
common to the both subsystems, we have $\Xi_{\rm A}=\Xi_{\rm B}$,
which is consistent to the intensive nature of $\Xi$.
We also have $\beta_{\rm A}=\beta_{\rm B}$ in equilibrium states.
These qualities lead to
\begin{equation}
  {\cal F}(S_{\rm A};{\rm M}_{\rm A}, N_{\rm A})
  ={\cal F}(S_{\rm B};{\rm M}_{\rm B}, N_{\rm B}).
\label{F-det}
\end{equation}
From the special property that ${\cal F}(S_{\rm X};{\rm M}_{\rm X}, N_{\rm X})$ is
intensive and independent of $V_{\rm X}$, we can write 
${\cal F}(S_{\rm X};{\rm M}_{\rm X}, N_{\rm X})
=\bar {\cal F}(s_X;{\rm M}_{\rm X})$ with $s_X\equiv S_X/N_X$. 
Here, if ${\rm M}_{\rm A}={\rm M}_{\rm B}={\rm M}$,
(\ref{F-det}) becomes $\bar {\cal F}(s_{\rm A};{\rm M})
=\bar {\cal F}(s_{\rm B};{\rm M})$. 
Because this holds any $s_{\rm A}$ and $s_{\rm B}$, 
we conclude that $\bar {\cal F}(s;{\rm M} )= c(\rm M)$,
where the constant $c({\rm M})$ depends not on $s$ but on
the type of material M. 
Thus, ${\cal F}(S;{\rm M},N) = c({\rm M})$ holds generally. 


Further, considering a general case ${\rm M}_{\rm A} \not = {\rm M}_{\rm B}$
for (\ref{F-det}), we have $c({\rm M_{\rm A}})= c({\rm M_{\rm B}})$ 
for any ${\rm M}_{\rm A}$ and ${\rm M}_{\rm B}$.
That is, ${\cal F}=c_*$ is a universal constant
independent of the type of material. From $c_*=\beta^{-1}\Xi$, 
the universal constant $c_*$ has the same dimension as the action, 
which is known as the Planck constant $\hbar $. Thus, our framework
based on classical theory has led to the existence of the Planck constant. 
Then, we can write $\xi= \hbar \beta$ and ${\cal F}=\hbar$,  where a dimensionless
proportionality constant has been chosen to be unity without loss
of generality. 


Finally, (\ref{integral}) leads to $\Psi+E \Xi=  \hbar S +b \hbar N$,
where $b$ is a dimensionless constant. We thus conclude that the
thermodynamic entropy $S$ is uniquely characterized as the Noether invariant
associated with the transformation $t \to t+\eta \hbar \beta$ 
for thermodynamically consistent trajectories \cite{SM-d7}. 
This is the main result of the present Letter. 

{\em Concluding remarks.---} 
First of all, 
we do not have a physical explanation of 
the symmetry for the real time transformation $t \to t+\eta \hbar \beta$ yet. 
It is interesting to find some relation with the fact that 
the complex time $t+i \hbar \beta$ naturally appears in quantum 
dynamics with finite temperature.  
An important point here is that the symmetry is an emergent 
property  in thermodynamic behavior of macroscopic systems, 
which can build a new bridge between microscopic 
and macroscopic physics as follows.  
	
One fascinating approach is to generalize this formulation to 
perfect fluids for interacting particles or relativistic fields, 
which could provide a more clear view to the symmetry. 
By restricting the spacetime configurations to those consistent 
with a local Gibbs distribution
at any time, we can find a symmetry leading to the local conservation 
of the entropy as the Noether charge. 
It seems reasonable to conjecture that this symmetry is 
explicitly observed in action functionals for perfect fluids,
although the action functionals are not uniquely determined
so far \cite{Brown}. With regard to this point, we also
mention a symmetry property announced  
in Ref. \cite{adiabatic-hydrodynamics,deBoer}, which may have
some relevance with our theory.


Although our study was motivated by the  black hole entropy 
as the Noether charge \cite{Wald}, it is not clear yet how 
the present analysis is related to that. 
Nevertheless, the
symmetry for 
$t \to t+\eta \hbar \beta$  
may correspond to that for the Killing parameter translation 
$v \to v+\eta \hbar \beta_{\rm H}$, where $\beta_{\rm H}$ 
is the inverse Hawking temperature \cite{Wald}. 
It would be 
interesting to investigate  connection of our theory
with a real-time and micro-canonical approach to thermodynamics 
of gravitational systems \cite{Brown2}. 


Finally, we have studied the invariant property of the
entropy in  quasi-static processes. More important is the  
non-decreasing property of entropy for general time dependent operations. 
If an  initial phase space point is sampled 
according to the equilibrium ensemble, this property can 
be proved \cite{Lennard, Jarzynski,Tasaki}. It is a challenging problem
to combine the symmetry property with the second law of thermodynamics,  
where the notion of thermodynamically consistent trajectories could be useful. 




The authors thank A. Dhar, A. Dechant, M. Hongo, M. Hotta, 
C. Jarzynski, A. Kundu, R. Loganayagam, C. Maes, and 
H. Tasaki for helpful discussions. 
This work was initiated when one of the authors (SS) visited 
Raman Research Institute (RRI).  He thanks the hospitality of RRI
and International Centre for Theoretical Sciences.  The present 
study was supported by KAKENHI (Nos. 25103002 and 26610115), 
and by the JSPS Core-to-Core program 
``Non-equilibrium dynamics of soft-matter and information''.


\onecolumngrid

\def\theequation{S\arabic{equation}}
\makeatletter
\@addtoreset{equation}{section}
\makeatother

\setcounter{equation}{0}

\newpage
\section{Supplemental Material}

\subsection{Mechanical description}
In order to consider the system concretely, 
we give an example of a Lagrangian.
Let $\bv{r}_i$ be the position of the $i$-th particle.  We 
assume the Lagrangian for $q=(\bv{r}_i)_{i=1}^N$ as 
\begin{equation}
L(q,\dot q, \alpha)=\sum_{i=1}^N \frac{m}{2}
\left( \der{\bv{r}_i}{t} \right)^2 - 
\sum_{i<j} U_{\rm int}(|\bv{r}_i-\bv{r}_j|) 
-\sum_{i=1}^N U_{\rm wall}(\bv{r}_i,\alpha),
\end{equation}
where $m$ is the mass of particles, $U_{\rm int}(r)$ is the 
interaction potential between two particles, and 
$U_{\rm wall}(\bv{r},\alpha) $ is the wall potential
confining particles. Explicitly, we define the bulk
region ${\cal D}=\{ (x,y,z)| 
0 \le x \le L_x,  0 \le y \le L_y,  0 \le z \le L_z \}$,
and we assume that $U_{\rm wall}(\bv{r},\alpha)=0 $ for 
$\bv{r} \in {\cal D}$ and $U_{\rm wall}(\bv{r},\alpha)=k(d/d_0)^2 $ for $\bv{r} \not \in {\cal D}$. 
Here $d_0$ is a positive constant that characterizes the width 
of the wall region, $d $ is the distance to ${\cal D}$ 
from $\bv{r} \not \in {\cal D}$, and $k$ is a positive constant. 
In order to represent the control by a piston, 
we fix  $L_y$ and $L_z$ to be $L$,
and set $L_x= \alpha/L^2$. Then, $\alpha$ is the volume 
of the bulk region, and the time dependence of $\alpha$ 
corresponds to the change in $L_x$ which is 
caused by the piston. 


When we consider a mechanical description 
on the basis of  a Lagrangian, we implicitly assume 
that the system is isolated from the other dynamical 
degrees of freedom. However, it seems impossible to 
justify this assumption for  experiments. 
For example, we may consider the above wall 
potential $U_{\rm wall}(\bv{r},\alpha)$ 
approximately as an effective one-body potential determined 
from the interaction between the system particles and the atoms 
constituting the wall. 
If the approximation were idealized, 
the dynamical degrees of freedom of the wall 
would not influence the motion of the particles. 
In experiments using adiabatic walls, however, 
the error in this approximation is not well-controlled, 
while the energy of the system is conserved within a measurement time. 
Thus, in general, 
trajectories realized in experiments are {\it not} given
by solution trajectories of the isolated system. 
Keeping this in mind, nevertheless, we study the isolated 
system as one idealization of the system. 

\subsection{Statistical mechanics}

We derive the relation (12) in the main text.
From  the definition 
\begin{equation}
\Omega(E,\alpha)\equiv \int d \Gamma \theta(E-H(\Gamma,\alpha)),
\end{equation}
we have 
\begin{eqnarray}
\pder{\Omega(E, \alpha)}{\alpha}
&=&   - \int d\Gamma \delta(E-H(\Gamma,\alpha)) \pder{H}{\alpha} \\
&=&   - \Sigma(E,\alpha) \bra \pder{H}{\alpha} \ket_{E,\alpha}^{\rm mc},
\end{eqnarray}
where we have used (10) in the main text. Then, from (11) in the main text,
we obtain
\begin{eqnarray}
\pder{S(E,\alpha)}{\alpha} &=& 
\pder{\log \Omega(E, \alpha)}{\alpha}  \nonumber \\
&=&  - \frac{\Sigma}{\Omega}\bra \pder{H}{\alpha} \ket_{E,\alpha}^{\rm mc} 
\nonumber  \\
&=&  - \beta \bra \pder{H}{\alpha} \ket_{E,\alpha}^{\rm mc} ,
\label{s-formula}
\end{eqnarray}
where we have used  the definition of $\beta$ given by (13) 
in the main text. 
Finally, by using 
\begin{equation}
\pder{\Omega}{E}=\int d\Gamma \delta (E-H(\Gamma,\alpha))=\Sigma(E,\alpha), \nonumber
\end{equation}
we also obtain 
\begin{eqnarray}
\beta(E,\alpha) &\equiv& \frac{\Sigma(E,\alpha)}{\Omega(E,\alpha)} 
\nonumber \\
                &= &  \pder{\log \Omega(E,\alpha)}{E} 
\nonumber \\
&= &  \pder{S(E,\alpha)}{E}.
\label{s-der-b} 
\end{eqnarray}
Thus, the relations (\ref{s-formula}) and (\ref{s-der-b}) 
mean (12) in the main text.

\subsection{Adiabatic theorem}


In this section, we review the adiabatic theorem. 
We consider the time evolution of 
$S_*(t) \equiv S(H(\Gamma_*(t),\alpha(t)),\alpha(t))$ 
along a solution trajectory $\Gamma_*(t)$ 
in the quasi-static
limit, where $\Gamma_*(t)$ satisfies 
the Hamiltonian equation
\begin{equation}
\dot q  =  \pder{H}{p},~~~\dot p  =  -\pder{H}{q}.
\end{equation}
Then, the adiabatic condition (16) holds. 
We start with (17) for $\Gamma_*(t)$: 
\begin{equation}
S_*(t_{\rm f})- S_*(t_{\rm i})
=  \int_{t_{\rm i}}^{t_{\rm f}} dt \beta_* \dot \alpha   
\left[ 
\left. \pder{H}{\alpha} \right\vert_* 
- \bra \pder{H}{\alpha} \ket_{E_*(t),\alpha(t)}^{\rm mc}
\right].
\label{adiabatic-theorem}
\end{equation}
The adiabatic theorem
claims that $S_*(t_{\rm f})= S_*(t_{\rm i})$ for almost all solution trajectories 
in the quasi-static limit. 
That is, the right-hand side of (\ref{adiabatic-theorem})
becomes zero in the quasi-static limit.


First, we explain a physical picture of the theorem.
We set $t_k=k \Delta +t_{\rm i}$ with $k=0, 1, \cdots, K$, 
where $t_K=t_{\rm f}.$ We choose $\epsilon$
satisfying $\epsilon \Delta \ll 1$ for a given $\Delta$. 
The key claim here is that there exists 
$\Delta $ such that 
\begin{equation}
 \frac{1}{\Delta}\int_{t_k}^{t_{k+1}} dt \left. \pder{H}{\alpha} \right\vert_*
=\bra \pder{H}{\alpha} \ket_{E_{*k},\alpha_k}^{\rm mc}
+ O( \epsilon \Delta )
\label{ergodic}
\end{equation}
for almost all solution trajectories, where $E_{k}=E(t_k) $
and $\alpha_k=\alpha(t_k)$. 
First, from $\alpha_{k+1} -\alpha_k 
= \int_{t_k}^{t_{k+1}} dt d \bar\alpha(\epsilon t)/dt$, we have 
$\alpha_{k+1} -\alpha_k = O(\epsilon \Delta)$. Then, by integrating 
the energy balance equation (15) 
during the time interval $[t_k, t_{k+1}]$
along a solution trajectory, which satisfies (16), 
we obtain $(E_{k+1} -E_k )_* = O( \ep \Delta)$, where we have used 
$\partial {H}/\partial \alpha =O(1)$ in the limit $ \epsilon \to 0$.
Thus, we may assume that solution trajectories are in the 
same energy surface during the time interval $[t_k, t_{k+1}]$ 
with ignoring $O( \epsilon \Delta) $ contribution. Now, 
if a phase space point at time $t_k$ is selected 
according to the micro-canonical ensemble, the probability that 
the value of  $\partial H/\partial \alpha$ is deviated from the 
typical value that is equal to the expectation value 
$\bra \partial H/\partial \alpha \ket_{E_{*k},\alpha_k}^{\rm mc}$ 
by a distance larger than some positive value, is exponentially 
small as a function of $N$. 
However, a phase space point at $t=t_k$ may become non-typical
by the influence of the operation $\alpha$ or we may select a non-typical
point with our special intention. 
Even for these cases, $\partial H/\partial \alpha$ 
approaches the typical value within a relaxation time 
$t_{\rm R}$ for almost all solution trajectories, because phase space 
points that take the typical value dominate the energy surface. 
Although there are still exceptional 
phase space points that do not exhibit the typical relaxation  behavior,
the probability of finding such phase space points 
is expected to be extremely small. Ignoring these exceptional trajectories, 
we choose $\Delta$ satisfying 
$\Delta \gg t_{\rm R}$ so that (\ref{ergodic}) holds. 
Finally, summing (\ref{ergodic}) for each step $k$, we have 
\begin{equation}
\int_{t_{\rm i}}^{t_{\rm f}} 
dt \left. \beta_* \dot \alpha \pder{H}{\alpha}\right\vert_*
= \sum_{k=0}^K  \beta_{*k} \dot \alpha \Delta 
\bra \pder{H}{\alpha} \ket_{E_{*k},\alpha_k}^{\rm mc}
+ O(  K \Delta^2 \epsilon^2 ),
\label{ergodic-2}
\end{equation}
where we have used $\dot \alpha =O(\epsilon)$ and $\beta=O(1)$. 
By considering the limit $ \epsilon\Delta \to 0$, $K \to \infty$,
and $t_{\rm R}^{-1}\Delta \to \infty$ with $K \epsilon \Delta 
=\tau_{\rm f}-\tau_{\rm i}$ fixed, we obtain 
\begin{equation}
\lim_{\epsilon \to 0} 
\int_{\tau_{\rm i}/\epsilon}^{\tau_{\rm f}/\epsilon} dt \beta_* \dot \alpha   
 \left[ 
\left. \pder{H}{\alpha} \right\vert_* 
- \bra \pder{H}{\alpha} \ket_{E_*(t),\alpha(t)}^{\rm mc}
\right]=0.
\label{at-result}
\end{equation}
Together with (\ref{adiabatic-theorem}), thus this gives an informal  proof of the adiabatic theorem


Next, we give a more formal explanation of the adiabatic theorem 
by following the method used in Ref. [15]. We define 
\begin{equation}
X(\Gamma,\alpha) \equiv 
\pder{H}{\alpha}- \bra \pder{H}{\alpha} \ket_{H(\Gamma,\alpha),\alpha}^{\rm mc}.
\label{Adef}
\end{equation}
Suppose that there exists a bounded function $\varphi(\Gamma,\alpha)$ 
satisfying 
\begin{equation}
X = 
\pder{\varphi}{q}\pder{H( \Gamma, \alpha)}{p}
-
\pder{\varphi}{p}\pder{H( \Gamma, \alpha)}{q} 
\label{assum}
\end{equation}
for almost all $\Gamma$. 
Then, since the right hand side is evaluated as 
\begin{eqnarray}
\left[ \der{\varphi}{t}- 
\der{\alpha}{t} \pder{\varphi}{ \alpha} 
\right]_* 
\end{eqnarray}
along a solution trajectory, we have
\begin{equation}
\int_{t_{\rm i}}^{t_{\rm f}} dt X(\Gamma_*(t),\alpha(t))
= 
\varphi_*(t_{\rm f})-\varphi_*(t_{\rm i})
-\int_{\tau_{\rm i}}^{\tau_{\rm f}}d\tau 
\left. 
\der{\bar \alpha}{\tau} 
\pder{\varphi}{\alpha}
\right\vert_* .
\end{equation}
By defining 
\begin{equation}
g(\tau)
\equiv \beta(\bar E(\tau),\bar \alpha(\tau)) 
\der{ \bar \alpha(\tau) }{\tau},
\end{equation}
and setting $C_0 =\sup_\tau |g(\tau)|$, 
$C_1 = |\varphi_*(t_{\rm f})-\varphi_*(t_{\rm i})|$,
and $C_2 = \left \vert \int_{\tau_{\rm i}}^{\tau_{\rm f}}d\tau 
({d \bar \alpha}/{d \tau})({\partial\varphi}/\partial {\alpha})_* 
\right\vert $,
we obtain 
\begin{equation}
\left\vert 
\epsilon \int_{\tau_{\rm i}/\epsilon}^{\tau_{\rm f}/\epsilon}dt 
g(\epsilon t) X(\Gamma_*(t),\alpha(t))  
\right\vert 
\le  \epsilon C_0 (C_1+C_2).
\end{equation}
Since $C_0$, $C_1$ and $C_2$ are independent of $\epsilon$, 
the adiabatic theorem holds in the quasi-static limit.
Thus, we have only to show that there exists $\varphi$ 
that satisfies (\ref{assum}).

Let us interpret (\ref{assum}) as a linear partial differential 
equation $X  = {\cal L}\varphi$ for  $\varphi$.
If ${\cal L}^{-1}$ exists, $\varphi$ is given as 
${\cal L}^{-1} X$. 
However, ${\cal L}^{-1}$ does not exist, 
since it is obvious that ${\cal L} f(H(\Gamma))=0$ for any function $f(E)$. 
From the assumption that 
there are no conserved quantities other than the energy, 
we may postulate that there are no other functions $f'$ 
such that ${\cal L} f'=0$. 
For any phase space functions
$g_1$ and $g_2$ in an appropriate function space, we define 
${\cal L}^\dagger$ as 
\begin{equation}
\int d\Gamma g_1(\Gamma) {\cal L} g_2(\Gamma)=
\int d\Gamma ({\cal L}^\dagger g_1(\Gamma) ) g_2(\Gamma).
\end{equation}
Because ${\cal L}^\dagger=- {\cal L}$, ${\cal L}^\dagger f(H(\Gamma))=0$.
Thus, the solution to $X  = {\cal L}\varphi$ exists 
when the solvability condition
\begin{equation}
\int d \Gamma f(H(\Gamma,\alpha)) X(\Gamma,\alpha) =0
\label{solv}
\end{equation}
holds. 
See the next paragraph for the explanation of the solvability 
condition. 
When the solvability condition (\ref{solv}) is satisfied  
for any $f$, we can say that there exists $\varphi$
to  $X  = {\cal L}\varphi$.
We here simplify the condition (\ref{solv}).
By substituting 
\begin{equation}
\int dE \delta (H(\Gamma,\alpha)-E) =1
\end{equation}
into (\ref{solv}), we have 
\begin{equation}
\int d \Gamma  
\int dE \delta(H(\Gamma,\alpha)-E)
f(H(\Gamma,\alpha)) X(\Gamma,\alpha)=0,
\label{solv-2}
\end{equation}
which is written as 
\begin{equation}
\int dE f(E) \Sigma(E,\alpha)\bra X \ket^{\rm mc}_{E,\alpha}=0.
\label{solv-3}
\end{equation}
Thus, the solvability condition (\ref{solv}) becomes
$\bra X  \ket^{\rm mc}_{E,\alpha} =0$
for any $E$ and $\alpha$. 
This is satisfied for $X(\Gamma,\alpha)$ defined by (\ref{Adef}). 
Thus, since there exists $\varphi$ that satisfies (\ref{assum}), 
we have reached the adiabatic theorem. 


Finally, in order to have a self-contained argument, we here  review 
the solvability condition for a linear algebraic 
equation for $\bv{x}$ in an $n$-dimensional vector space. We study 
\begin{equation}
M \bv{x} = \bv{b},
\label{linear-a}
\end{equation}
where $M$ is an $n \times n$ matrix, and $\bv{b}$ is a constant vector.
When $M^{-1}$ exists, the solution is obtained as 
\begin{equation}
\bv{x}= M^{-1} \bv{b}.
\end{equation}
However, when there exists $\bv{y}_0$ such $M \bv{y}_0=0$, 
$M^{-1}$ does not exist. We assume that there are no other 
zero-eigenvectors. In this case, whether the solution 
to (\ref{linear-a}) exists or not depends on $\bv{b}$. 
Concretely, let $M^\dagger$ be the adjoint matrix defined by
\begin{equation}
(\bv{u}, M \bv{v})=(M^\dagger\bv{u},  \bv{v})
\end{equation}
for any vectors $\bv{u}$ and $\bv{v}$, 
where $(\ , \ )$ denotes the standard inner product in the vector space.
Let $\bv{z}_0$ be  the left zero-eigenvector defined by $M^\dagger \bv{z}_0
=0$. Then, when $(\bv{z}_0, \bv{b}) \not = 0$, there is no solution to 
(\ref{linear-a}). When  
\begin{equation}
(\bv{z}_0, \bv{b}) = 0,
\label{s-con}
\end{equation}
we have an infinite number of solutions
\begin{equation}
\bv{x}=M_{\rm ps}^{-1} \bv{b} +\chi \bv{y}_0,
\end{equation}
where $\chi$ is an arbitrary number and 
$M_{\rm ps}^{-1}$ is the pseudo-inverse matrix of $M$  
such that $M_{\rm ps}^{-1} M \bv{u}= M M_{\rm ps}^{-1} \bv{u} =\bv{u}$
for any $\bv{u} \not \in {\rm Ker}(M)$. (\ref{s-con}) is referred to as 
the solvability condition. 


It should be noted that the argument presented above is not
mathematically rigorous. Even if a systematic approximation
of $\varphi$ in (\ref{assum}) as a finite dimensional
vector $\bv{x}$ in (\ref{linear-a}) is found, the limit to $\varphi$
from $\bv{x}$  is not obvious at all. For example,
Ref. [15] proved that there exists a smooth function
$\varphi$ such that the ${\rm L}^2$ norm of $X  -{\cal L}\varphi$ is less
than any positive $\epsilon $. This weak statement implies that there
is no smooth function $\varphi$ satisfying (\ref{assum}). 
From a different viewpoint, it was pointed out that the solution
$\varphi$ in (\ref{assum}) is not a standard function but should be
a distribution [19]. We thank Christopher Jarzyski for these
particular comments. 

\subsection{Thermodynamically consistent trajectories}

In this section, we demonstrate some examples of thermodynamically
consistent trajectories that satisfy the condition (14). First, we check
that almost all solution trajectories satisfy (14). The statement is
basically equivalent to the adiabatic theorem. Indeed, instead of 
(\ref{ergodic-2}), we can write
\begin{equation}
\int_{t_{\rm i}}^{t_{\rm f}} 
dt \left.  \dot \alpha \pder{H}{\alpha}\right\vert_*
= \sum_{k=0}^K  \dot \alpha \Delta 
\bra \pder{H}{\alpha} \ket_{E_{*k},\alpha_k}^{\rm mc}
+ O(  K \Delta^2 \epsilon^2 )
\label{ad-ad}
  \end{equation}
By considering the limit $ \epsilon\Delta \to 0$, $K \to \infty$,
and $t_{\rm R}^{-1}\Delta \to \infty$ with $K \epsilon \Delta 
=\tau_{\rm f}-\tau_{\rm i}$ fixed, we obtain 
\begin{equation}
\lim_{\epsilon \to 0} 
\int_{\tau_{\rm i}/\epsilon}^{\tau_{\rm f}/\epsilon} dt \dot \alpha   
 \left[ 
\left. \pder{H}{\alpha} \right\vert_* 
- \bra \pder{H}{\alpha} \ket_{E_*(t),\alpha(t)}^{\rm mc}
\right]=0.
\label{ad-ad2}
\end{equation}
This means that the solution trajectories satisfy (14).

Next, we explicitly show that (14) is satisfied for non-solution trajectories
consistent with quasi-static isothermal processes.
Concretely,  suppose that we have a trajectory $\hat q_{\rm tot}$ 
for the total system consisting of a system and a heat bath, 
whose trajectories are given by $\hat q$ and $\hat q_{\rm bath}$, 
respectively.  
Because the total system is ideally isolated, 
a solution trajectory for the total Lagrangian
$\hat q_{*\rm tot}$ is realized. 
Then, a trajectory $\hat q$ which is obtained by projecting $\hat q_{*\rm tot}$ to the system is not a solution trajectory 
for the system Lagrangian, because of the interaction with $\hat q_{\rm bath}$. 
Nevertheless, when a quasi-static operation is performed 
to the system, the energy of the system $E$ is determined 
from the condition $\beta(E,\alpha)={\rm const}$ and the trajectory 
$\hat q$ satisfies the condition (14).  We shall prove this claim. 


Let $\Gamma_{\rm tot}=(\Gamma, \Gamma_{\rm B})$ be a phase space
point of the composite system, 
where  $\Gamma \in {\mathbb{R}}^{6N}$ and 
$\Gamma_{\rm B}  \in {\mathbb{R}}^{6N_{\rm B}}$ 
are for the system and heat bath, respectively, with $N_{\rm B} \gg N \gg 1$. 
We assume 
\begin{equation}
H_{\rm tot}(\Gamma_{\rm tot},\alpha)
=H(\Gamma,\alpha)+H_{\rm B}(\Gamma_B)+H_{\rm int}(\Gamma,\Gamma_B)
\simeq H(\Gamma,\alpha)+H_{\rm B}(\Gamma_B),
\end{equation}
where $H_{\rm int}$ can be ignored in the evaluation 
of the statistical average. Specifically, we consider the
statistical average of the thermodynamic quantity 
\begin{equation}
Y(\Gamma)=\pder{H(\Gamma,\alpha)}{\alpha}
\label{Y-def}
\end{equation}
with respect to the micro-canonical ensemble of the total
system, that is,
\begin{equation}
\bra Y  \ket^{\rm mc:tot}_{E_{\rm tot},\alpha}
\equiv \frac{1}{\Sigma_{\rm tot}(E_{\rm tot},\alpha)}
\int d \Gamma_{\rm tot} 
\delta(H_{\rm tot}(\Gamma_{\rm tot},\alpha)-E_{\rm tot})
Y(\Gamma).
\end{equation}
We can define $S_{\rm tot}$, $\Omega_{\rm tot}$, $\Sigma_{\rm tot}$ 
and $\beta_{\rm tot}$ from $H_{\rm tot}$ similarly, and also 
define $S_{\rm B}$, $\Omega_{\rm B}$, $\Sigma_{\rm B}$ 
and $\beta_{\rm B}$ from  $H_{\rm B}$. 
Note here that for a solution trajectory 
$\Gamma_{\rm tot*}$ of the total system 
\begin{equation}
\left. \frac{dH_{\rm tot}}{dt}\right|_{\Gamma_{\rm tot*}} 
= \left. \pder{H}{\alpha} \right|_{\Gamma_{\rm tot*}} \dot \alpha 
\label{dot_H_tot}
\end{equation}
holds, and because $S_{\rm tot}(E_{\rm tot},\alpha)=S_{\rm B}(E_{\rm tot})+
O(N)$, we have
\begin{equation}
\beta_{\rm tot}= \pder{S_{\rm B}}{E_{\rm tot}} +O\left( \frac{N}{N_{\rm B}}
\right),
\end{equation}
which leads to
\begin{eqnarray}
\frac{d}{dt}\beta_{\rm tot}(E_{*\rm tot},\alpha)
&=& \left. \frac{\partial^2 S_{\rm B}}{\partial E_{\rm tot}^2}
\right|_* \dot E_{*\rm tot} +O\left( \frac{N}{N_{\rm B}}
\right) \nonumber \\
&=&
\left. \frac{\partial^2 S_{\rm B}}{\partial E_{\rm tot}^2}
\right|_* 
\left. \pder{H}{\alpha} \right|_{\Gamma_{\rm tot*}}\dot \alpha
+O\left( \frac{N}{N_{\rm B}} \right) \nonumber \\
&=&
O\left( \frac{N}{N_{\rm B}} \right).
\end{eqnarray}
We thus assume that 
$\beta_{\rm tot}$ is a constant value $\tilde \beta$
in the quasi-static limit. 
Then, as we will show later, there exists $E$ such that 
\begin{eqnarray}
\tilde \beta &=& \beta(E,\alpha) ,
\label{beta-eq} \\  
\bra Y  \ket^{\rm mc:tot}_{E_{\rm tot},\alpha}
&=& 
\bra Y  \ket^{\rm c}_{\tilde \beta,\alpha},
\label{eq-ens-1}  \\
\bra Y  \ket^{\rm c}_{\tilde \beta,\alpha}
&=& \bra Y  \ket^{\rm mc}_{E,\alpha},
\label{eq-ens}
\end{eqnarray}
where 
\begin{eqnarray}
\bra Y  \ket^{\rm c}_{\tilde\beta,\alpha}
&\equiv&
\frac{1}{Z(\tilde\beta,\alpha)}\int d \Gamma e^{-\tilde\beta H(\Gamma,\alpha)}Y(\Gamma) 
\label{can-def}
\nonumber \\
\bra Y  \ket^{\rm mc}_{E,\alpha}
&\equiv& \frac{1}{\Sigma(E,\alpha)}
\int d \Gamma \delta(H(\Gamma,\alpha)-E)Y(\Gamma),
\end{eqnarray}
with the normalization constant $Z$ given by
\begin{equation}
Z(\tilde\beta,\alpha)=\int d\Gamma e^{-\tilde\beta H(\Gamma,\alpha)}.
\label{Z-def}
\end{equation}
The equality (\ref{eq-ens}) is called the equivalence of ensembles.

When (\ref{beta-eq}), (\ref{eq-ens-1}) and (\ref{eq-ens}) hold, 
we can show that (14) is satisfied for the system trajectory 
$\hat q$, as follows. 
First, by applying (\ref{ad-ad2}) to a solution trajectory 
$\Gamma_{\rm tot*}$ of the total system, 
and employing (\ref{dot_H_tot}), 
we obtain 
\begin{equation}
\int_{t_{\rm i}}^{t_{\rm f}} dt \dot \alpha   
\left[ 
\left. \pder{H}{\alpha} \right\vert_{\Gamma_{\rm tot*}}
- \bra \pder{H}{\alpha} \ket_{E_{\rm tot*}(t),\alpha(t)}^{\rm mc:tot}
\right]=0
\label{at-result-2}
\end{equation}
in the quasi-static limit. 
Here, the first term in the 
integral in (\ref{at-result-2}) is evaluated at the system 
trajectory $\Gamma$ obtained by projecting $\Gamma_{\rm tot*}$ 
to the system, and  the second term is rewritten by using 
(\ref{eq-ens-1}) and (\ref{eq-ens}). We then reach (14) 
for the system trajectory $\hat q$ which is not the 
solution but consistent with quasi-static isothermal processes. 

\subsection{Equivalence of ensembles}

In this section, we derive (\ref{beta-eq}), (\ref{eq-ens-1})
and (\ref{eq-ens}). As a preliminary, we note the asymptotic behavior 
\begin{equation}
\frac{\Omega(E,\alpha)}{N!} 
= \exp\left[ N \omega \left( \frac{E}{N}, \frac{\alpha}{N} \right) 
+o(N) \right]
\end{equation}
for short-range interacting particles systems, where $\alpha$ is assumed
to be an extensive parameter such as the volume. This gives 
\begin{equation}
S(E,\alpha) = \log \frac{\Sigma(E,\alpha)}{N!} +o(N).
\label{omega-sigma}
\end{equation}
Similarly, we have
\begin{equation}
S_{\rm B}(E_{\rm B}) = \log \frac{\Sigma_{\rm B}(E_{\rm B})}{N_{\rm B}!} +o(N_{\rm B}).
\end{equation}
For any variable $A(\Gamma)$, we can write 
\begin{equation}
\bra A  \ket^{\rm mc:tot}_{E_{\rm tot},\alpha}
= \frac{1}{\Sigma_{\rm tot}(E_{\rm tot},\alpha)}
\int d \Gamma A(\Gamma) \Sigma_{\rm B}(E_{\rm tot}-H(\Gamma,\alpha)),
\label{1st}
\end{equation}
where we have used 
\begin{equation}
\Sigma_{B}(E_{\rm B})=\int d\Gamma_{\rm B} 
\delta (H_{\rm B}(\Gamma_{\rm B})-E_{\rm B}).
\end{equation}
We calculate 
\begin{eqnarray}
&& \log \left [
\frac{\Sigma_{\rm B}(E_{\rm tot}-H(\Gamma,\alpha)) }{N_{\rm B}!} 
\right ] \nonumber \\
= &&
S_{\rm B}(E_{\rm tot}-H(\Gamma,\alpha))+o(N_{\rm B}) \nonumber \\
= &&
S_{\rm B}(E_{\rm tot})
- \beta'  H(\Gamma,\alpha)
+O\left (\frac{N}{N_{\rm B}} \right) +o(N_{\rm B})
\end{eqnarray}
with 
\begin{equation}
\beta' \equiv 
\left. \der{\log \Sigma_{\rm B}(E)}{E} \right|_{E=E_{\rm tot}}.
\end{equation}
From $S_{\rm tot}(E_{\rm tot},\alpha)= S_{\rm B}(E_{\rm tot})+ O(N)$,
we find that $\beta' = \tilde \beta +O(N/N_{\rm B})$. By ignoring the
term $O(N/N_{\rm B})$, we obtain $\beta' = \tilde \beta$. 
The expression (\ref{1st}) is now rewritten as
\begin{equation}
\bra A  \ket^{\rm mc:tot}_{E_{\rm tot},\alpha}
= \frac{\Sigma_{\rm B}(E_{\rm tot})}{\Sigma_{\rm tot}(E_{\rm tot},\alpha)}
\int d \Gamma A(\Gamma) e^{- \tilde \beta H(\Gamma,\alpha)+O(N/N_{\rm B})}. 
\end{equation}
Hereafter, we ignore the term $O(N/N_{\rm B})$. 
By setting $A=1$, we find 
\begin{equation}
 \frac{\Sigma_{\rm B}(E_{\rm tot})}{\Sigma_{\rm tot}(E_{\rm tot},\alpha)}
=\frac{1}{Z(\tilde \beta,\alpha)}.
\end{equation}
We thus obtain
\begin{equation}
\bra A  \ket^{\rm mc:tot}_{E_{\rm tot},\alpha}
= \bra A \ket^{\rm c}_{\tilde \beta,\alpha}.
\end{equation}
By setting $A=Y$, we thus have obtained (\ref{eq-ens-1}).

Next, we consider $\bra Y \ket^{\rm c}_{\tilde \beta,\alpha}$.
By using  (\ref{Y-def}), (\ref{can-def}) and (\ref{Z-def}), we have
\begin{equation}
\pder{\log Z(\tilde \beta,\alpha)}{\alpha}
=-\tilde \beta \bra Y \ket^{\rm c}_{\tilde \beta,\alpha}.
\label{Z-der}
\end{equation}
We rewrite $\log Z $ as follows. By substituting 
\begin{equation}
\int dE' \delta(H(\Gamma,\alpha)-E')=1
\end{equation}
into (\ref{Z-def}), we have 
\begin{eqnarray}
\frac{Z(\tilde \beta,\alpha)}{N!}
&=&  
\frac{1}{N!}
\int d\Gamma \int dE' \delta(H(\Gamma,\alpha)-E') e^{-\tilde \beta H(\Gamma,\alpha)} 
\nonumber \\
&=&  
\int dE' e^{-\tilde \beta E'} \frac{\Sigma(E',\alpha) }{N!}
\nonumber \\
&=& 
\int dE' e^{-\tilde \beta E'+S(E',\alpha) +o(N)},
\end{eqnarray}
where we have used (\ref{omega-sigma}). 
We ignore the term $o(N)$. The saddle point estimation leads to
\begin{eqnarray}
\frac{Z(\tilde \beta,\alpha)}{N!}
&=& 
\exp\left[ - \inf_{E'}[\tilde\beta E'-S(E',\alpha)] +O(\log N) \right]
\nonumber \\
&=&
\exp\left[-[\tilde\beta E(\tilde\beta,\alpha)-S(E(\tilde\beta,\alpha),\alpha)] +O(\log N)\right],
\end{eqnarray}
where $E(\tilde \beta,\alpha)$ is the minimizer of 
$\tilde \beta E'-S(E',\alpha)$. Therefore, 
we obtain
\begin{equation}
\tilde \beta= \pderf{S}{E}{\alpha},
\end{equation}
which is identified as (\ref{beta-eq}).
We also have 
\begin{equation}
\log \frac{Z(\tilde\beta, \alpha)}{N!} 
=-[\tilde\beta E(\tilde\beta,\alpha)-S(E(\tilde\beta,\alpha),\alpha)].
\end{equation}
By combining this with (\ref{Z-der}) and employing (\ref{s-formula}),  we obtain
\begin{eqnarray}
\bra Y \ket^{\rm c}_{\tilde\beta,\alpha}
&=& \pderf{E}{\alpha }{\tilde\beta}
-\left. \tilde\beta^{-1}\pderf{S}{\alpha}{E} \right\vert_{E=E(\tilde\beta,\alpha)}
-\left. \tilde\beta^{-1}\pderf{S}{E}{\alpha} \right\vert_{E=E(\tilde\beta,\alpha)}
 \pderf{E}{\alpha }{\tilde\beta}
\nonumber \\
&=&-\tilde\beta^{-1} \left.  \pderf{S}{\alpha }{E} \right\vert_{E=E(\tilde\beta,\alpha)}
\nonumber \\
&=& \bra Y \ket^{\rm mc}_{E(\tilde\beta,\alpha),\alpha} .
\end{eqnarray}
We have arrived at (\ref{eq-ens}). 


\subsection{Derivation of (21)}

We here derive (21) in the main text. 
To do that, we first check that 
\begin{equation}
\lim_{\epsilon \to 0} \int_{\tau_{\rm i}}^{\tau_{\rm f}} d\tau 
\frac{d \alpha}{d\tau}  \Xi 
\left[ 
\pder{H}{\alpha} 
-
\bra \pder{H}{\alpha} \ket^{\rm mc}_{\bar E(\tau), \bar \alpha(\tau)}
\right]=0
\label{19-goal}
\end{equation}
holds for thermodynamically consistent trajectories. 
Noting $\Xi(\tau)=\Xi(\bar E(\tau),\bar \alpha(\tau))$ and
setting $\tau_k=\tau_{\rm i}+(\tau_{\rm f}-\tau_{\rm i})k/K$,  
the left-hand side of (\ref{19-goal}) is estimated as 
\begin{equation}
  \sum_{k=1}^K \Xi(\tau_{k})
  \int_{\tau_{k-1}}^{\tau_{k}} d\tau   \frac{d\bar \alpha}{d\tau}
  \left[\pder{H}{\alpha}-\bra \pder{H}{\alpha}
    \ket_{\bar E(\tau), \bar\alpha(\tau)}^{\rm mc}\right] +O(1/K)
\label{19-1}
\end{equation}
for large $K$ limit. Thus, taking the limit $\epsilon \to 0$, 
the left-hand side of (\ref{19-goal}) becomes $O(1/K)$ due to (14).
After that, taking the limit $K \to \infty$, then we obtain (\ref{19-goal}). 
Now, by using (\ref{19-goal}) and (20) for $\xi=\Xi$ and $\psi=\Psi$, 
we have (21).


\subsection{Derivation of $\Xi$}

In this section, we solve (23) in the main text, which is 
expressed by
\begin{equation}
\pder{}{\alpha}\left(\Xi \beta^{-1} \pderf{S}{E}{\alpha}\right)_E
- \pder{}{E}
\left( \Xi \beta^{-1} \pderf{S}{\alpha}{E}  \right)_\alpha =0,
\label{target}
\end{equation}
where we have used (24) in the main text. 
Then, by setting $\Phi=\Xi \beta^{-1}$, we find that the Jacobian
determinant $|\partial(\Phi,S)/\partial(\alpha,E)|$ is zero. 
We take one curve $\Phi(E,\alpha)=\phi_0$, where $\phi_0$ is
a constant. Then, the tangent vector at any point $(E,\alpha)$ 
on the curve are perpendicular to $(\partial_E \Phi, \partial_\alpha \Phi)$,   
which is proportional to $(\partial_E S, \partial_\alpha S)$ because 
$|\partial(\Phi,S)/\partial(\alpha,E)|=0$. Thus, there exists a curve 
$S(E,\alpha)=s_0$ with a constant $s_0$  whose tangent vector 
at any point $(E,\alpha)$ on the curve is proportional to that of
$\Phi(E,\alpha)=\phi_0$. This implies that the curve 
$S(E,\alpha)=s_0$ is identical to $\Phi(E,\alpha)=\phi_0$.
We thus obtain $\Phi={\cal F}(S)$, where ${\cal F}$ is an arbitrary
function. 

We can also derive this result without geometrical consideration.
We express (\ref{target}) as 
\begin{equation}
\pderf{\Phi}{\alpha}{E} \beta
+
\pderf{\Phi}{E}{\alpha}\pderf{E}{\alpha}{S}\beta=0,
\label{phi-ident}
\end{equation}
where we have used the relation (12) and 
\begin{equation}
\pderf{S}{\alpha}{E}=-\pderf{E}{\alpha}{S} \pderf{S}{E}{\alpha}.
\end{equation}
Furthermore, by taking the derivative of $\Phi(E(S,\alpha),\alpha)$
with respect to $\alpha$, we find that 
\begin{equation}
\pderf{\Phi}{\alpha}{S}
=
\pderf{\Phi}{\alpha}{E} 
+
\pderf{\Phi}{E}{\alpha}\pderf{E}{\alpha}{S}.
\label{phi-der}
\end{equation}
From (\ref{phi-ident}) and (\ref{phi-der}), 
we have 
\begin{equation}
\pderf{\Phi}{\alpha}{S}=0.
\end{equation}
This means that 
$\Phi={\cal F}(S)$, where ${\cal F}$ is an arbitrary
function. 


\subsection{Special case}

One may consider the special case where the parameter $\alpha$ 
does not depend on time, $\alpha_0={\rm const}$.  
The equation (20) in the main text is 
then satisfied if we find $\xi$ and $\psi$ such that 
\begin{equation}
\der{\psi}{t} +  E \dot\xi =0.
\label{eq-target}
\end{equation}
Suppose that $\xi = \Xi(E(q,\dot q,\alpha_0),\alpha_0)$ and 
$\psi = \Psi(E(q,\dot q,\alpha_0),\alpha_0)$ satisfy (\ref{eq-target}).
For any such function $\Xi(E,\alpha_0)$, we obtain the Noether invariant
\begin{equation}
\Psi+E \Xi= \int^E dE' \Xi(E',\alpha_0). 
\label{special}
\end{equation}
This formula means that 
there exists a transformation leading to the conservation
of any function of energy through the Noether theorem, 
which includes the special case where the energy $E$ itself is 
conserved for the uniform time translation $\Xi={\rm const}$.
In particular, by choosing $\Xi=\hbar \beta$, we obtain
\begin{eqnarray}
\Psi+E \Xi &=&  \hbar \int^E dE' \beta(E',\alpha_0) \nonumber \\
           &=& \hbar S(E,\alpha_0)+S_0(\alpha_0).
\end{eqnarray}
This is consistent with our result. However, in contrast to 
the argument in the main text, this consideration cannot 
lead to the unique characterization of the entropy as 
the Noether invariant.  Thus, the consideration of quasi-static
processes with the time-dependent parameter $\alpha (t)$ is 
inevitable to obtain our main result.


\begin{thebibliography}{10}
\bibitem{fn:Noether}
The original work \cite{Noether} discussed 
the case where $\psi=0$, which was generalized to $\psi\neq0$ by Ref. \cite{Bessel-Hagen}.
Ref. \cite{Trautmen} investigated the generalization precisely 
and discussed the property that 
a solution is mapped to another one by such a transformation.
This property was referred to as a dynamical symmetry in 
Refs. \cite{Prince} and \cite{SIAM}.
\bibitem{fn:symmetry}
The symmetry is mathematically expressed as the invariant 
property of the contact form in the tangent bundle 
\cite{Prince, Crampin, SIAM}. This symmetry is called 
the $d\theta$-symmetry \cite{SIAM} or the Cartan symmetry \cite{Prince}.
\bibitem{S_hbar}
One might notice the absence 
of the factor $(2\pi \hbar)^{3N}$ in (\ref{S-formula}), 
but this definition is sufficient when we are not concerned 
with the additive constant of the entropy.
\end{thebibliography}
\end{document}